# On multi-scale percolation behaviour of the effective conductivity for the lattice model

W. Olchawa[a], R. Wiśniowski[b,1], D. Frączek[c,2], R. Piasecki[d,*]

[a,b,d] *Institute of Physics, University of Opole, Oleska 48, 45-052 Opole, Poland*
[c] *Department of Materials Physics, Opole University of Technology, Katowicka 48, 45-061 Opole, Poland*

HIGHLIGHTS

- We explore the multi-scale percolation behaviour in a few lattice models.
- A change in the size of the basic cluster alters the value of the percolation threshold.
- In the extended three-phase model, double-threshold routes appear on the conductivity surface.

ABSTRACT

Macroscopic properties of heterogeneous media are frequently modelled by regular lattice models, which are based on a relatively small basic cluster of lattice sites. Here, we extend one of such models to any cluster's size $k \times k$. We also explore its modified form. The focus is on the percolation behaviour of the effective conductivity of random two- and three-phase systems. We consider only the influence of geometrical features of local configurations at different length scales $k$. At scales accessible numerically, we find that an increase in the size of the basic cluster leads to characteristic displacements of the percolation threshold. We argue that the behaviour is typical of materials, whose conductivity is dominated by a few linear, percolation-like, conducting paths. Such a system can be effectively treated as one-dimensional medium. We also develop a simplified model that permits of an analysis at any scale. It is worth mentioning that the latter approach keeps the same thresholds predicted by the former one. We also briefly discuss a three-phase system, where the double-thresholds paths appear on model surfaces.

(Some figures in this article are in colour only in the electronic version)

## 1. Introduction

One of the simplest techniques embodying the essential physics of macroscopic properties of random heterogeneous materials is the network extension of effective medium approximation (EMA) [1-3]. A lot of data relating to experimental macroscopic conductivity can be quantitatively fitted by a general effective medium (GEM) equation, which combines most aspects of percolation and effective medium theories [4]. From the theoretical point of view, the percolation phenomena are still an interesting area of active research, for instance, concerning a new effective medium theory that considers a global tunnelling network model of conductor-insulator composites [5], where the percolation and the tunnelling regimes are treated on equal footing. Interestingly, the percolation transition itself can be an 'explosive' in nature [6]. The recent development of a scaling theory of this exotic transition provides the full set of scaling functions, critical exponents and explains its continuous nature and unusual properties [7].

[*] Corresponding author. Tel.: +48 77 452 7285; fax: +48 77 452 7290.
E-mail addresses: wolch@uni.opole.pl (W. Olchawa), ryszard.wisniowski@gmail.com (R. Wiśniowski), d.fraczek@po.opole.pl (D. Frączek), piaser@uni.opole.pl (R. Piasecki).
[1] R. Wiśniowski is a Ph.D. student at Institute of Physics, University of Opole.
[2] D. Frączek is a recipient of a Ph.D. scholarship under a project funded by the European Social Fund.

Lattice approaches are particularly suited for computing transport properties in solutions of complex fluids. In the recent paper, Hattori *et al*. have proposed a specific but tractable lattice model considering, within EMA, the percolation behaviour of electrical conductivity of interacting droplets in micro-emulsion solutions [8]. For these purposes, they make use of the basic cluster composed of only four unit cells and despite of this gross simplification their results are in a qualitative agreement with the experimental data [8]. However, the connection between the percolation threshold and particles interaction is not a subject of the present paper. On the other hand, such an approach is of interest to us in the specific context of the effective properties of anisotropic multi-phase random heterogeneous media; see e.g. [9].

Here we investigate how the results in the Hattori *et al*. approach depend on the size of the basic cluster. We focus on the influence of *geometrical* features of local configurations in the classes specified by the model *without* particles interaction but at *different* length scales $k$. Thus, we confine the analysis of the $k$-multiscale percolation behaviour to a high-temperature limit of the model [8]. As long as the developed general formula is employed, we consider local configurations in square clusters until their size (the scale $k$) is accessible on a personal computer. Furthermore, some simplified formulas based on presumed criterions turn out to be easy to compute analytically at any scale. This allows checking a suggestion of Ref. [8] related to possible changes in the slope of the effective conductivity curve of a two-phase system, together with the increase in the basic cluster's size. Since the hallmark of percolation in two-phase systems composed of low and high-conductivity components is a sharp increase in the effective electrical conductivity at the critical concentration, we apply a similar definition to multi-phase systems with dominating components of the lowest and highest conductivity.

It is worthy of remark that in [8] the droplets are treated as finite-size objects, i.e. unit square cells centred on the lattice sites what leads to the site percolation. In this paper, the mono-sized particles are represented by pixels. This would enable us to include the needed information directly from a digitized image of a real sample. This is a significant point that the effective conductivity and percolation itself can be influenced by the size of mono-grains as well as by the distribution of grain size in a material. The question was studied by mean of the EMA using a coarse square lattice with the model bonds, on example of binary disordered media with complex distributions of grain size [10].

This paper is organised as follows: In Section 2, we consider a general cluster of size $l \times k$ instead of $2 \times 2$ basic clusters discussed in [8]. We introduce the extended formula for the probability of appearance of allowed classes of configurations of non-interacting particles at any length scale $k$. The combinatorial computations are limited numerically to small scales $k \leq 12$, yet. Therefore, we also develop a simplified model and present a formula that permits to do analytical calculations at any scale $k$. In Section 3, we analyse the role of dimensionality in modification of the extended model, as well as its simplified form. In Section 4, we exemplify the extension to a three-phase system composed of low, medium and high conductivity components. Finally, we briefly summarize the results and make general conclusions in Section 5.

## 2. Extended Z4-model accounting for all local conductivities

Let us consider a scale extension of the two-dimensional lattice EMA-based model used by Hattori *et al*. [8]. For a unit lattice-distance, square cells 1×1 are centred on the lattice sites. Each of them can be occupied by a particle, here represented by a black pixel, of the *H*-phase with a given electrical conductivity $\sigma_H >> \sigma_M$. For the *M*-phase, its $\sigma_M$-conductivity is attributed to an empty cell represented here by a white pixel. In this paper, we restrict ourselves to the simplest case of a random system of non-interacting particles.

However, the main assumption is still held, i.e. the fluctuations of electrical potential in the orthogonal direction to the macroscopic electric field are neglected [8]. (Without any loss of the generality, the electrical field is directed along *x*-axis.) Inspired by this assumption, we focus on investigation of a possible multi-scale behaviour of electrical percolation in such a model. Thus, we consider a general case of an elementary $l \times k$ cluster instead of $2 \times 2$ basic one applied earlier [8]. At this stage, our main task is to obtain the probability of appearance of a class of configurations, each of the local conductivity $\sigma_x(\{n_m\})$ computed accordingly to the formula

$$\sigma_x(\{n_m\}) = \frac{k}{l} \sum_{m=0}^{k} \frac{n_m}{m/\sigma_H + (k-m)/\sigma_M} \; . \tag{2.1}$$

Here $l$ and $k$ denote the number of rows and columns, while $n_m$ describes the number of rows with exactly $m$ black pixels representing $H$-phase. It should be stressed that the quantity $n_m \in \{n_0, n_1, \ldots, n_k\}$ is the basic random variable, whereas the other variables are expressed by a mean of the $n_m$. The second remark relates to the very specific feature of the considered models, i.e. the local conductivity is not the weighted (by local volume concentrations) average of conductivities of the two phases. We remind that the following obvious conditions should also be fulfilled:

$$n_0 + n_1 + \cdots + n_k = l \tag{2.2a}$$

and

$$n_1 + 2n_2 + \cdots + kn_k = N_H \; , \tag{2.2b}$$

where $0 \leq N_H \leq kl$ is a number of black pixels of the considered cluster. It should also be noticed that the local conductivity given by Eq. (2.1) is not dependent on the permutations of pixels in any row and on the permutations of rows, either.

Now, we are in a position to give the needed formula for the probability of the appearance of a class of configurations having a local conductivity $\sigma_x$,

$$P(\{n_m\}) = \frac{l!}{n_0! n_1! \ldots n_k!} p_0^{n_0} p_1^{n_1} \cdots p_k^{n_k} \; , \tag{2.3a}$$

where the probability of a row occupation by $m$ particles of $H$-phase reads

$$p_m = \frac{k!}{(k-m)! m!} \phi_H^m (1 - \phi_H)^{k-m} \; . \tag{2.3b}$$

Here, $\phi_H$ denotes the global volume fraction of phase $H$. It is clear that Eqs. (2.3a, b) represent the appropriate multinomial and binomial Bernoulli distributions.

To illustrate the possible classes of configurations, i.e. those with an identical conductivity $\sigma_x$, we present in Fig. 1 half of them for the case of $3 \times 3$ clusters. The 'symmetrical' remaining part can be obtained by making the following replacements: white phase $\leftrightarrow$ black one, $\sigma_H \leftrightarrow \sigma_M$ and $\phi_H \to 1 - \phi_H$.

| Class | Local conductivity | Probability |
|---|---|---|
| 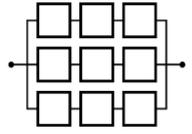 | $\sigma_M$ | $(1-\phi_H)^9$ |
| 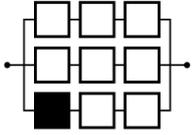 | $\dfrac{2\sigma_M}{3} + \dfrac{\sigma_M \sigma_H}{\sigma_M + 2\sigma_H}$ | $9\phi_H(1-\phi_H)^8$ |
| 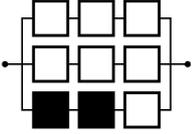 | $\dfrac{2\sigma_M}{3} + \dfrac{\sigma_M \sigma_H}{2\sigma_M + \sigma_H}$ | $9\phi_H^2(1-\phi_H)^7$ |
| 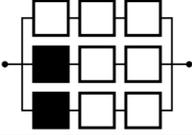 | $\dfrac{\sigma_M}{3} + \dfrac{2\sigma_M \sigma_H}{\sigma_M + 2\sigma_H}$ | $27\phi_H^2(1-\phi_H)^7$ |
| 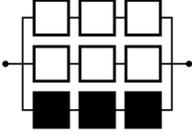 | $\dfrac{2\sigma_M + \sigma_H}{3}$ | $3\phi_H^3(1-\phi_H)^6$ |
| 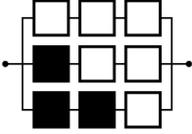 | $\dfrac{\sigma_M}{3} + \dfrac{\sigma_M \sigma_H}{\sigma_M + 2\sigma_H} + \dfrac{\sigma_M \sigma_H}{2\sigma_M + \sigma_H}$ | $54\phi_H^3(1-\phi_H)^6$ |
| 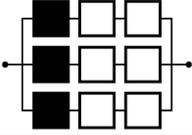 | $\dfrac{3\sigma_M \sigma_H}{\sigma_M + 2\sigma_H}$ | $27\phi_H^3(1-\phi_H)^6$ |
| 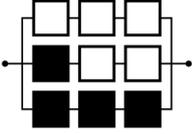 | $\dfrac{\sigma_M + \sigma_H}{3} + \dfrac{\sigma_M \sigma_H}{\sigma_M + 2\sigma_H}$ | $18\phi_H^4(1-\phi_H)^5$ |
| 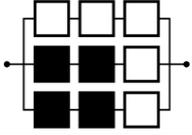 | $\dfrac{\sigma_M}{3} + \dfrac{2\sigma_M \sigma_H}{2\sigma_M + \sigma_H}$ | $27\phi_H^4(1-\phi_H)^5$ |
| 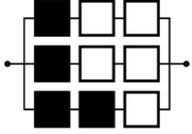 | $\dfrac{2\sigma_M \sigma_H}{\sigma_M + 2\sigma_H} + \dfrac{\sigma_M \sigma_H}{2\sigma_M + \sigma_H}$ | $81\phi_H^4(1-\phi_H)^5$ |

**Fig. 1.** Representation of the different possible classes of configurations in two dimensions at length scale $k = 3$. The local conductivities $\sigma_x$ are given in the second column. The corresponding probabilities are presented in the last column. The 'symmetrical' remaining part is not shown. It can be obtained by making the following replacements: white phase $\leftrightarrow$ black one, $\sigma_H \leftrightarrow \sigma_M$ and $\phi_H \to 1 - \phi_H$.

Following the idea of the effective medium, a hypothetical regular network is considered, where the conductance of the resistor on each bond is of the same value. Within the Bruggeman approach [1, 2], the effective conductivity $\sigma^*$ is given by

$$\sum_{\{n_m\}} P(\{n_m\}) \frac{\sigma_x(\{n_m\}) - \sigma^*}{\sigma_x(\{n_m\}) + (z/2 - 1)\sigma^*} = 0, \quad (2.4)$$

where $z$ denotes the coordination number of the EMA-network. The proper term can be equivalently rewritten as $(z/2 - 1) \equiv (d - 1)$, where $d$ is the dimension of space.

In Sections 2 and 3, the numerical results relate to a two-phase system with components of higher conductivity, $\sigma_H = 10^6$, and a moderate one, $\sigma_M = 1$ in a.u. Now, for simplicity, $l = k$ and the square side of length $k$ is the measure of length scale. In Fig. 2a, the $k$-multi-scale dependence of the total effective conductivity $\sigma^*(k; Z4)$ as a function of volume fraction $\phi_H$ is demonstrated. For chosen scales, $k = 2, 3, \ldots 12$, one can observe displacements of the location of percolation threshold toward higher values of the critical concentration $(\phi_H)_c$. (We limit scale $k = 12$ to the maximal because of the large number 2704156 of the distinct classes, whereas the entire number of local configurations is about $2.2 \times 10^{43}$).

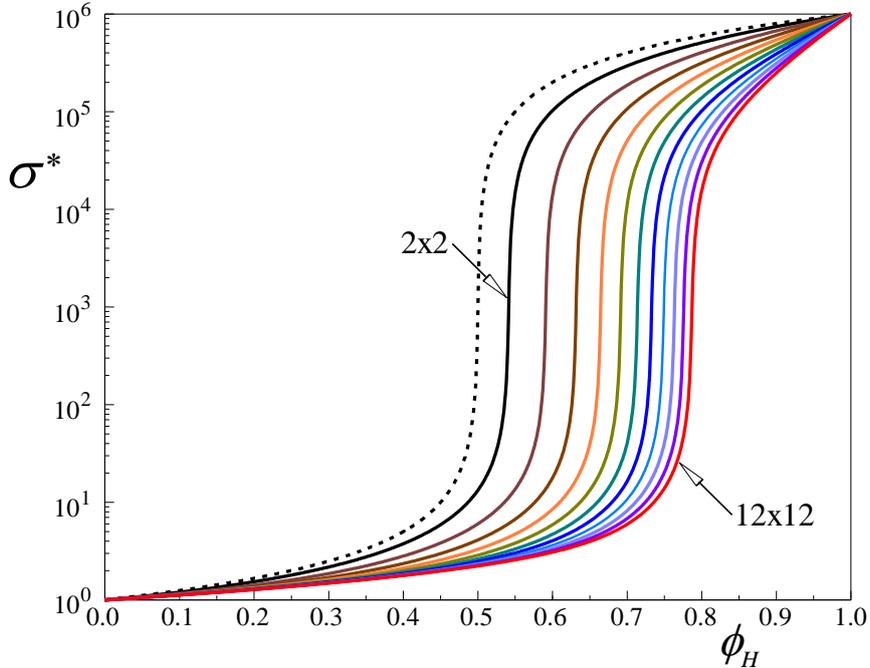

**Fig. 2a.** Evolution of the effective conductivity $\sigma^*(k; Z4)$ as a function of the volume fraction $\phi_H$ of highly conducting $H$-phase for $k \times k$ local clusters at length scales, $k = 2, 3 \ldots 12$. These scales are accessible numerically. The conductivity of $H$-phase and $M$-phase is $10^6$ and 1 in a.u., respectively. The shifting of the percolation threshold is clearly seen along $k$-scale. For completeness, the case of the simplest $1 \times 1$ clusters is also presented (the dashed line).

This kind of behaviour seems to be not supporting the remark given by Hattori *et al.* [8] (top line on page 36). According to the cited opinion: "As a consequence (i.e. of the using of a cluster of larger sizes – our comment) the change in conductivity versus $\Phi$ would likely be smoother (i.e. the shifting is not expected – our comment)".

### 2.1 Simplified Z4s-model with a reduced number of local conductivities

The non-standard evolution of $\sigma^*(k;Z4)$ can be justified making use of the corresponding simplified model. It employs only the $\sigma_H$ and $\sigma_M$ conductivities attributed to the two classes of cluster configurations. Roughly speaking, among all the configurations there appear those we call *H*-class, i.e. with at least one row fully occupied by the phase *H*. Consequently, they have local conductivities of order $\sigma_H$, while the remaining configurations denoted as *M*-class have local conductivities of order $\sigma_M$; compare Eq. (2.1). Next, for both classes, the following approximation can be used:

$$\sigma_x(\{n_m\}) \approx \frac{n_k}{l} \sigma_H \approx \sigma_H \qquad \text{for} \quad n_k > 0 \qquad (2.1.1)$$

and

$$\sigma_x(\{n_m\}) \approx \sigma_M \frac{k}{l} \sum_{m=0}^{k-1} \frac{n_m}{(k-m)} \approx \sigma_M \qquad \text{for} \quad n_k = 0. \qquad (2.1.2)$$

Generally, for a given fraction $\phi_H$, the probability $P_M$ of the appearance of a configuration belonging to *M*-class depends on the length scales $k$ and $l$,

$$P_M(k,l) \equiv P(n_k = 0) = \sum_{\{n_0, n_1, \ldots, n_{k-1}\}} P(\{n_0, n_1, \ldots, n_{k-1}, 0\}) = (1 - \phi_H^k)^l. \qquad (2.1.3)$$

For the complementary *H*-class, the related probability $P_H$ simply equals

$$P_H(k,l) \equiv P(n_k > 0) = 1 - P(n_k = 0) = 1 - (1 - \phi_H^k)^l. \qquad (2.1.4)$$

Consequently, the general formula (2.4) describing the effective conductivity $\sigma^*$ can be simplified to

$$P_H(k,l)\frac{\sigma_H - \sigma^*}{\sigma_H + (z/2-1)\sigma^*} + P_M(k,l)\frac{\sigma_M - \sigma^*}{\sigma_M + (z/2-1)\sigma^*} = 0 . \qquad (2.1.5)$$

The question arises: How does the increase in the length scale $k$ for a square basic cluster, i.e. with $l = k$, alter the probability of each of the two classes $P_H(k)$ and $P_M(k)$ given by Eqs. (2.1.3-4)?

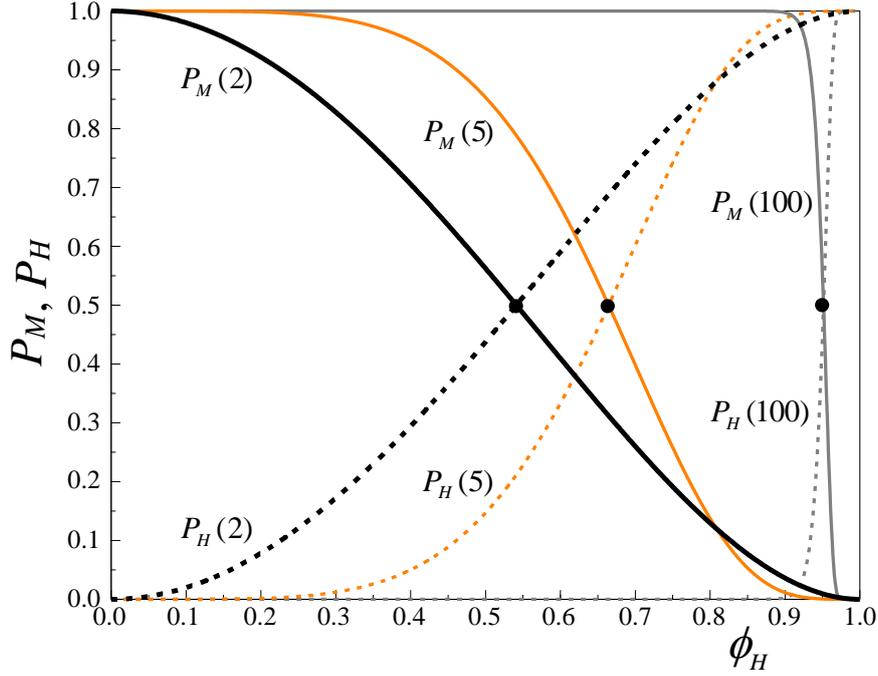

**Fig.2b.** The probability of the appearance of a class of configurations with the local conductivity $\sigma_x$ of order $10^6$ (the $P_H$ for $H$-class) and 1 (the $P_M$ for $M$-class), see Eqs. (2.1.1-4). For the simplified Z4s-model, the $\phi_H$-coordinate of the so-called balance point roughly corresponds to the critical volume fraction. At exemplary scales, $k = 2, 5$ and $100$, the balance points are marked by the black dots.

The conclusion inferred below for the chosen scales $k = 2, 5$ and $100$ and presented in Fig. 2b can be extended to any $k$ for a pair of the probabilities $P_H(k)$ and $P_M(k)$. Here, the first exemplary pair relates to the probability $P_M(2)$, the thick solid black line, and the $P_H(2)$, the thick dashed black one. A similar description applies to the probabilities of the next pairs with the related thin lines, orange and grey online. For each pair, the curves intersect at the so-called balance point marked by the bold dot. The points indicate approximately the pertinent critical concentration $(\phi_H(k))_c$ or equivalently, the relevant percolation threshold. Analytically, the positions of the percolation thresholds are given by

$$P_H(k) = P_M(k) \equiv 1/2 \quad \Rightarrow \quad (\phi_H(k))_c = (1-(1/2)^{1/k})^{1/k} . \qquad (2.1.6)$$

For example, here we have $(\phi_H(2))_c < (\phi_H(5))_c$. Correspondingly, along the length scale the locations of the successive balance points are clearly shifted on the right. This way one can understand the effect of shifting of the percolation threshold in Fig. 2a for the $\sigma^*(k; Z4)$-curves. The further illustration of the balance points for $k = 2$ and 5 will be given in Section 3.1 in Fig. 3c, where the collection of models is depicted.

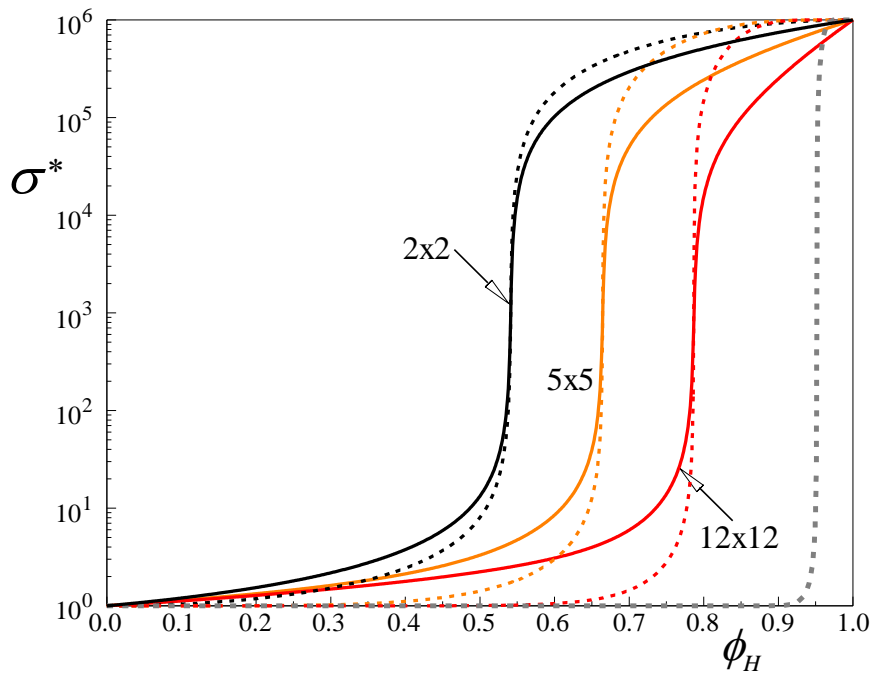

**Fig. 2c.** A comparison between the effective conductivity $\sigma^*(k; Z4)$, the solid lines, and $\sigma^*(k; Z4s)$, the dashed lines, at the chosen scales. Despite the different conductivity values (at fixed scale $k$) outside of the critical volume fraction, the positions of the percolation thresholds overlap; see the related curves for $k = 2, 5$ and 12. Notice, that the last presented case referring to scale $k = 100$, the thick grey line, is available only for the Z4s-model.

For completeness, in Fig. 2c, at given exemplary scales, we also present a comparison between the evolving effective conductivity $\sigma^*(k; Z4)$ described by Eq. (2.4), the solid lines, and the $\sigma^*(k; Z4s)$ given by Eq. (2.1.5), the dashed lines. It is interesting that the positions of the percolation threshold seem to overlap despite the differences between the $\sigma^*(k; Z4s)$-values and the $\sigma^*(k; Z4)$ ones in the neighbourhood of critical concentrations; compare the corresponding curves for $k = 2, 5$ and 12. The last presented case that refers to the scale $k = 100$, the thick grey line, is accessible numerically only for the Z4s-model. It should be underlined that within the framework of the Z4s-model, the threshold criterion $P_H = P_M \equiv \tfrac{1}{2}$ leads at any length scale $k$ to the effective conductivity

$$\sigma^*((\phi_H)_c) = \sqrt{\sigma_H \sigma_M} \ . \qquad (2.1.7)$$

This result deserves a short comment. In the literature, there is the well-known duality transformation of two-dimensional heterogeneous composites and the resulting exact result - similar to the above but for the volume fraction equal to ½ - for the effective conductivity of a two-phase system with a symmetric and isotropic distribution of components [11]. Here, considering elementary cell with $l = k = 1$, a two-phase system comprises topologically equivalent phases. According to Eqs. (2.1.3-4) and (2.1.6) we obtain $(\phi_H(k=1))_c = ½$ and as expected, the Keller-Dykhne formula [11] is reproduced.

## 2.2 Test of the numerical code

We decide to perform an additional test of the program code developed. We implement it to randomly generated two-phase patterns of the finite size 101×101 and of different volume concentrations $\phi_H$. Then, making use of a sliding sampling cell $k \times k$, we calculate experimental probabilities, $P(\{n_m\})_{exper}$. To get an effective experimental conductivity $\sigma^*_{exper}(k; Z4)$ as a function of the volume fraction $\phi_H$, we apply Eq. (2.4).

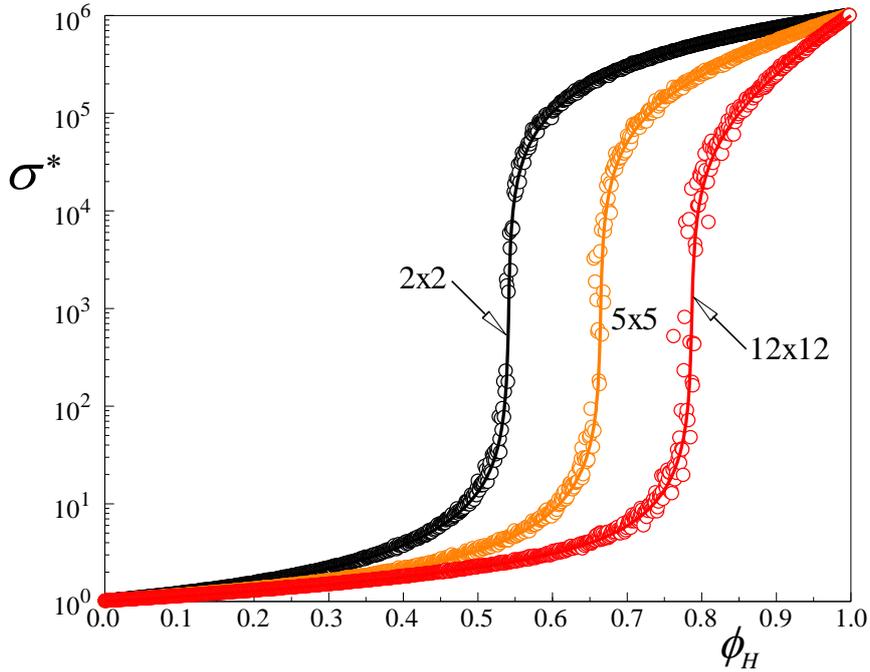

**Fig. 2d.** A comparison between the effective conductivity $\sigma^*(k; Z4)$, the solid lines, and its experimental counterparts $\sigma^*_{exper}(k; Z4)$, the open circles, obtained for one realization of the two-phase random 101×101 patterns at selected scales. Around the percolation thresholds, the fluctuations of experimental conductivity appear. This test confirms the correctness of the program code developed.

Fig. 2d shows the outcomes for the selected length scales $k = 2, 5$ and 12. The lines describe the same theoretical values of $\sigma^*(k; Z4)$ as in Fig. 2c, while the open circles correspond to the $\sigma^*_{\text{exper}}(k;Z4)$ calculated just for one realization of the corresponding random pattern. Each of the random patterns was created by a mean pseudo-random number generator [12]. One can observe that around the percolation thresholds, the larger the scale is, the greater the fluctuations of experimental conductivities appear. As expected, at the exemplary scales the values of theoretical effective conductivities are consistent with those obtained for the random patterns.

### 3.  Modified Z2-model accounting for all local conductivities

The results contained in Section 2 are obtained for $z = 4$ that corresponds to $d = 2$ used in Ref. [8]. It means that this dimension is equivalent to taking into account all directions on the isotropic square global lattice. This may cause some kind of internal inconsistency resulting in a consideration of local currents only in $x$-direction and simultaneously, global network currents in both directions: $x$ and y. Further investigation of locations of the percolation threshold needs some care devoted to the space dimension in the EMA approach; see Eq. (2.4).

It would be reasonable to investigate a global current allowed only in $x$-direction, i.e. in the same direction as it is assumed for any local configuration. This assumption suggests the usage of coordination number $z = 2$, so we have effectively a one-dimensional network. Such an approach seems to be another physically admissible option. Let us check its validity. One can imagine a supporting lattice model of a finite system containing $L$ lines, each occupied by $K$ resistors. Every resistor represents a local $\sigma_\alpha$-conductivity. For such a system, within the EMA approach, again the corresponding hypothetical network comprising only identical resistors $\sigma^*$ can be introduced. Now, the electrical potential $U$ applied along each of the network lines causes a total current

$$i^* = U \frac{L\sigma^*}{K} \tag{3.1}$$

Let one of the identical resistors be replaced by a resistor of local $\sigma_\alpha$-conductivity, where $\alpha \equiv \{n_m\}$. Recalling that the notation $n_m$ relates to the number of lines with exactly $m$ particles of the $H$-phase, the corresponding total current can be written as

$$i_\alpha = U \frac{1}{(K-1)/\sigma^* + 1/\sigma_\alpha} + U \frac{(L-1)\sigma^*}{K}. \tag{3.2}$$

Executing the above replacement with a probability denoted now as $P_\alpha$, we should remember that it is still given by Eq. (2.3a). Then, a fluctuation of the total current defined as $\Delta i_\alpha = i_\alpha - i^*$ can be treated as a random variable. It simply equals

$$\Delta i_\alpha = U\left[\frac{1}{(K-1)/\sigma^* + 1/\sigma_\alpha} - \frac{\sigma^*}{K}\right]. \tag{3.3}$$

It should be noticed that the parameter $L$ related to the number of rows (each of them being occupied by $K$ resistors) is missing in the above formula. Thus, the approach is not sensitive to the number of lattice rows.

Next, as required by the EMA, the expectation value of the fluctuation should be equal to zero. Thus, we put $<\Delta i_\alpha> = 0$ and from Eq. (3.3) we obtain a modified equation for the effective conductivity

$$\sum_\alpha P_\alpha \frac{\sigma_\alpha - \sigma^*}{\sigma_\alpha + \sigma^*/(K-1)} = 0. \tag{3.4}$$

Now, the limiting behaviour of the network can be evaluated for the infinite system, i.e. when $K \to \infty$. Equivalently, we may insert to Eq. (2.4) the coordination number $z=2$ (or equivalently $d=1$). This means that we get effectively a one-dimensional network within EMA. So, the Z2-model originates from this kind of an internal consistency: all local and global currents flow only along the x-direction.

As the simplest application of the one-dimensional EMA, the effective conductivity $\sigma^*$ for an infinite wire ($L=1$, $K \to \infty$) with any cluster of size $1 \times k$ equals

$$\sigma^* = \left\langle \frac{1}{\sigma_\alpha} \right\rangle^{-1} = \frac{1}{\phi/\sigma_H + (1-\phi)/\sigma_M}. \tag{3.5}$$

Interestingly, for this specific case of $1 \times k$-clusters, the multiscale behaviour becomes completely independent of the length scale. It is a clear result since only a series connection of resistors appears in this case.

Using now the limiting form ($K \to \infty$) of Eq. (3.4), we are in a position to recalculate the results obtained in Section 2. Fig. 3a shows the total effective conductivity $\sigma^*(k; Z2)$ as a function of volume fraction $\phi_H$ for different sizes of square cells $k \times k$.

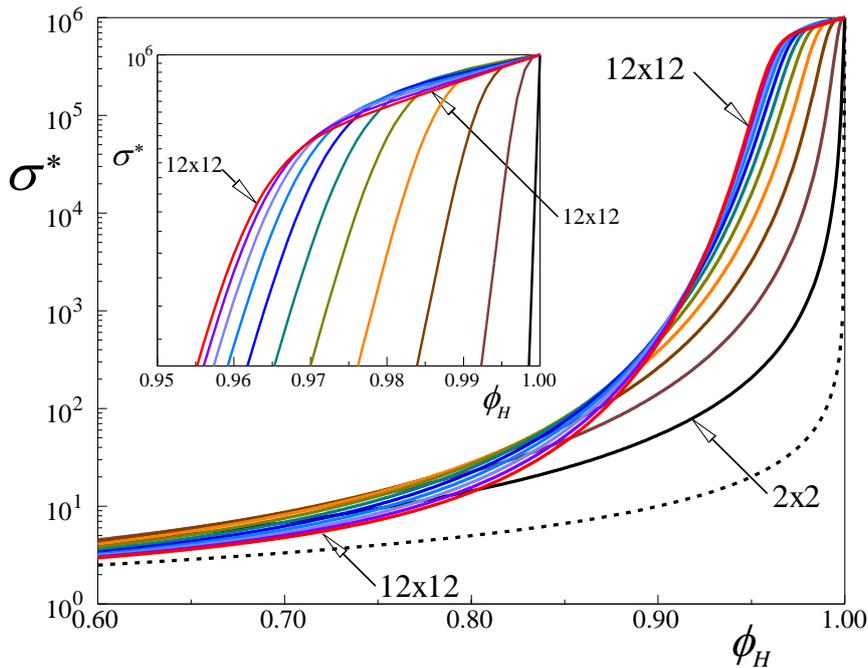

**Fig. 3a.** The same as in Fig. 2a but for the Z2-model. This time, along the length scales, $k = 2, 3 \ldots 12$, which are accessible numerically, we observe the reverse sequence of the conductivity curves for increasing scales. In contrast to the previous Z4-case any of two considered curves intersect twofold within the range $0 < \phi_H < 1$. For completeness, the case of the simplest 1×1 clusters is also presented, the dashed line; see Eq. (3.1.5). Additional details for the higher volume fraction can be seen in the inset.

For the same fixed length scales, $k = 2, 3 \ldots 12$, which are accessible numerically, now the percolation behaviour is different from that of earlier models and more complicated. For a higher volume fraction, more details can be seen in the inset of Fig. 3a. Careful observation indicates that every two $\sigma^*(k; Z2)$-curves among those considered, are intersecting two-fold within the range $0 < \phi_H < 1$. Now, a reverse sequence of the conductivity curves for the increasing scales, $k = 2, 3 \ldots 12$, can be seen within a restricted range of concentration. The further details referred to percolation behaviour will be given in the next section.

## 3.1 Simplified Z2s-model with reduced number of local conductivities

By analogy to Section 2.1, we apply again the idea of the simplified approach but now with $z = 2$. Let us denote it as the Z2s-model. We need to recall that this way permits of an evaluation of the effective conductivity at larger scales for $k > 12$. Applying Eq. (3.4) with the condition $(K \to \infty)$, the $\sigma^*(k; Z2s)$-conductivity is obtained as a function of volume fraction $\phi_H$. In Fig. 3b, we compare the evolution of the modified effective conductivity $\sigma^*(k; Z2)$, the solid lines, with its simplified counterparts $\sigma^*(k; Z2s)$, the dashed lines. For every $k$-pair of the solid and dashed line, except the single dashed curve referred to the length scale $k = 100$, one can observe rather a poor agreement between both effective conductivity values. Regardless of these differences, we propose quite a natural way to specify an approximate localization of the percolation threshold for the Z2s-model.

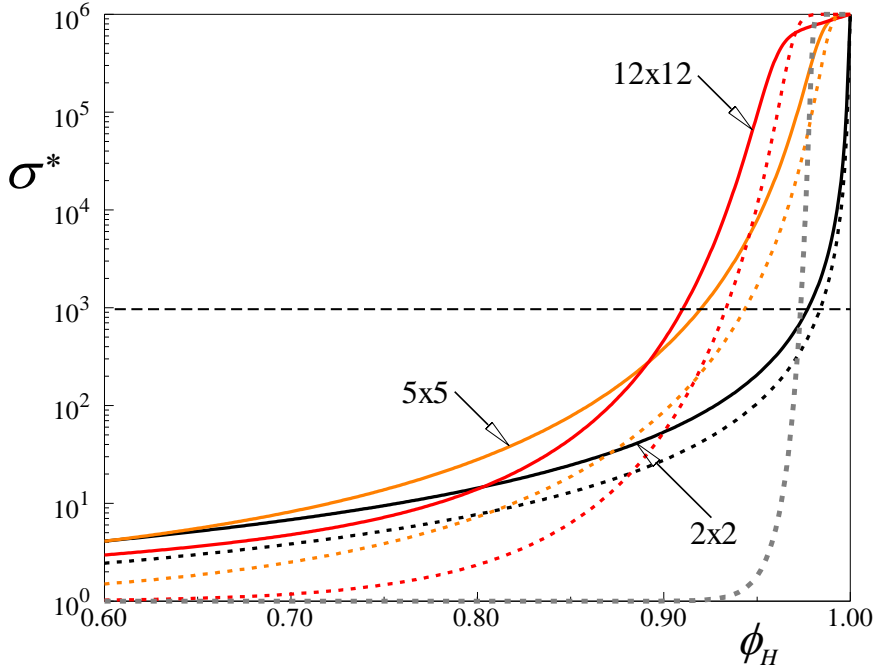

**Fig. 3b.** The same as in Fig. 2d but for the Z2-model, the solid lines, together with its simplified Z2s-version, the dashed lines. In contrast to the previous models the approximate percolation threshold is estimated by the intersection point of the conductivity level $\sigma^* = 10^3$ (the horizontal thin dashed line) determined according to the supposed criterion (see the text) with the corresponding effective conductivity line (solid for $\sigma^*(k; Z2)$ or dashed for $\sigma^*(k; Z2s)$).

Suppose that similarly to the Z4s-model already considered, at unknown yet critical volume fraction $(\phi_H)_c$, the effective conductivity should be very close to $\sigma^*((\phi_H)_c) = 10^3$ in a.u., when the same phase conductivities, the higher $\sigma_H = 10^6$ and moderate one $\sigma_M = 1$ are used. Now, using Eq. (2.1.5) with $P_H = 1 - P_M$ together with the above-presumed criterion, we derive a simple estimation of the $P_M$ probability

$$P_M = \frac{\sqrt{\sigma_H \sigma_M} - \sigma_M}{\sigma_H - \sigma_M} \approx \sqrt{\frac{\sigma_M}{\sigma_H}} = 10^{-3}. \tag{3.1.1}$$

This means that the probability of appearance of a cluster belonging to the lower conductivity *M*-class must be negligible to get a significant increase in the effective conductivity. Such behaviour is well understood because of the quantitative differences between the EMA for a square array (Z4s-model) employed in Section 2 and the effectively one-dimensional network used in this section. Indeed, the *H*-class and *M*-class configurations, having the high $\sigma_H$ local conductivity and the moderate $\sigma_M$ one, play a role of resistors connected in a series. Therefore, the main factor deciding on the total conductivity of any row (for the Z2s-model) is the absence of a resistor belonging to the *M*-class. Reversely, for the former Z4s-case, the so-called balance points involving the both classes are obviously the most important. Generalizing the Eq. (2.1.6) with the probability $P_M$ treated now as a parameter, the corresponding volume fraction can be expressed as

$$\phi_H(k;P_M) = (1-(P_M)^{1/k})^{1/k} . \tag{3.1.2}$$

Then, inserting the $P_M = 0.001$ in compliance with Eq. (3.1.1), we derive the needed critical volume fraction as the $\phi_H$-coordinate of the intersection point of a given $\sigma^*(k;Z2s)$-curve with the dashed line determined by the condition $\sigma^* = 10^3$. This is illustrated in Fig. 3b for some exemplary scales *k*.

Now, we are in a position to describe the significance of the curves referred to the fixed values of $P_M = $ **0.5**, 0.4, 0.3, 0.2, 0.1, 0.05, 0.01 and **0.001** in Fig. 3c. (Of course, equivalent values of $P_H = $ **0.5**, 0.6, 0.7, 0.8, 0.9, 0.95, 0.9 and **0.999** can be used as well.) The most important are the bold marked $P_M$-probabilities, **0.5** for the Z4s-model and **0.001** for the Z2s-model, which are of direct physical importance. The former (latter) parameter value corresponds to the monotonically increasing thick bottom curve with the marked balance points for $k = 2$ and 5 (the non-monotonic thick upper one – grey online). These curves relate to the *k*-multi-scale dependence of percolation threshold of the corresponding models. The minimum of the uppermost thick curve appearing at scale $k = 10$, originates the reverse shifting (compared to the lowermost thick curve) of the percolation threshold, i.e. when $k \leq 10$. The standard shifting on the right direction appears when $k > 10$.

To present the limiting convergence in both simplified models at $k \to \infty$, in the inset of Fig. 3c, the threshold locations as a function of the reciprocal of the length scale, $1/k$, are shown. The bottom (upper) filled circles correspond to Z4-model (Z2), while the bottom monotonic (upper non-monotonic) solid curves relate to Z4s-case (Z2s). In contrast to the approximation used in Z2s-model, the bottom curve corresponding to Z4s-case reproduces the locations of the threshold remarkably well. Interestingly, in the recent lattice model of randomly distributed overlapping squares (cubes in 3D) of any linear size in lattice units, the percolation of obstacles as well as void space was investigated in Ref. [13]. Employing the excluded volume approximation to discrete systems and using it to study the transition between continuous and discrete percolation, a remarkable agreement between the theory and numerical results was found. The authors illustrate, in a similar manner, that the percolation threshold of obstacles is a non-monotonic function of the obstacle size, while the percolation threshold of the void space is nearly a linear function of the reciprocal of the obstacle size; *cf*. Figs. 4-5 in [13].

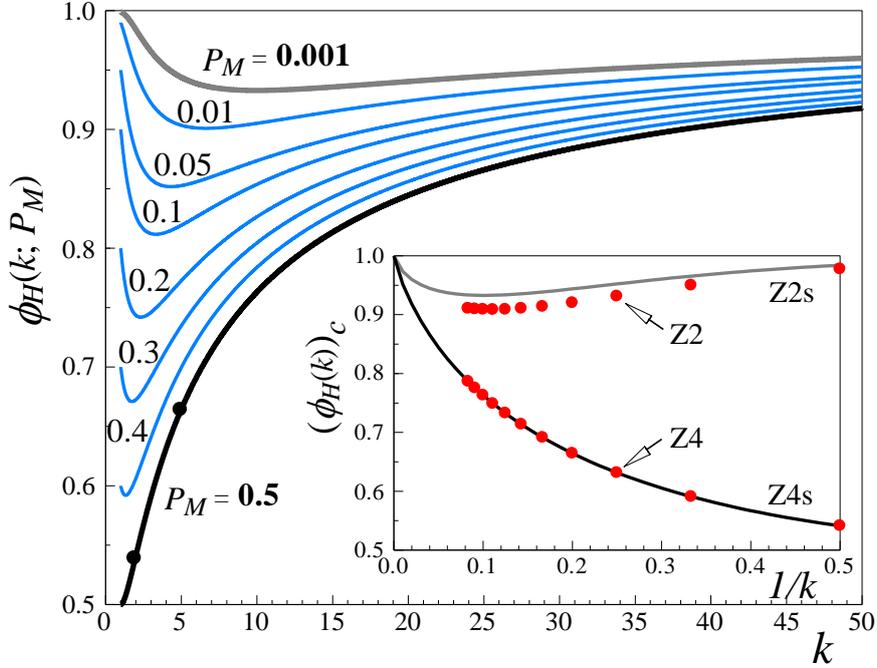

**Fig. 3c.** The parameterisation $P_M = $ **0.5**, 0.4, 0.3, 0.2, 0.1, 0.05, 0.01 and **0.001** for the collection of $\phi_H(k; P_M)$-curves, the solid lines; see Eq. (3.1.2). The bold marked $P_M$-probabilities are of direct physical significance. The lowermost monotonic thick black line (the uppermost non-monotonic thick grey one) describes $k$-multi-scale dependence of the critical volume fraction for the Z4s-model (Z2s), respectively. The minimum of top Z2s-curve at scale $k = 10$, reveals the reverse (compared to the bottom curve) shifting of the percolation threshold at scales $k \leq 10$. The common asymptotic behaviour is given by Eq. (3.1.3). In the inset, the percolation thresholds, as a function of the reciprocal of the length scale $1/k$, are shown.

On the other hand, the exemplary non-monotonic internal curves (blue on-line) in Fig. 3c, are not essential for an analysis of percolation behaviour within the presumed criterions. However, they still illustrate a kind of Z4s-Z2s crossover when the other $P_M$-parameter values are tested as a hypothetical criterion. Interestingly, for the formula (3.1.2), which describes all the $P_M$-curves, there is a simple asymptotic estimate irrespective of the value of $P_M$,

$$\phi_H(k; P_M) \approx 1 - \frac{\ln k}{k} \qquad \text{for } k \gg 1 \ . \tag{3.1.3}$$

Let us consider the border curves of the Z4s-model ($z = 4$) for the $P_M = $ **0.5** as well as of the Z2s-model ($z = 2$) for the $P_M = $ **0.001**. It is clear now that the significance of the parameter $z$ decreases gradually with the increase in the length scale $k$. A good example of such a behaviour is the nearly step-like $P_M$-line for $k = 100$, as it is shown in Fig. 2b.

## 4. Three-phase Z4 and Z2 models

There is another natural extension of the models considered in the earlier sections. It was stated in Ref. [9]: "The multi-component, multiphase materials are increasingly used in various fields, but analysis and investigation efforts are severely lagging behind". A good example is the structurally interesting porous manganite-insulator composite $La_{0.7}Ca_{0.3}MnO_3/Mn_3O_4 \equiv LCMO/Mn_3O_4$ [14]. Generally, in this practical three-dimensional medium, three different microstructures can be observed as the LCMO content decreases: (i) $Mn_3O_4$ islands in a LCMO matrix, (ii) a labyrinth pattern of the two phases, and (iii) LCMO islands in a $Mn_3O_4$ matrix; compare the electron micrographs of polished cross sections of the corresponding samples in Fig. 1 of [14]. The composite samples were considered as made up of a high-conductivity part (LCMO) and a low-conductivity part (Mn3O4 + air pores). The authors report temperature dependence of the zero-field resistivity of samples containing manganite volume fractions $f_{LCMO}$ as well as the $\Phi_c \sim 0.19$ experimental critical concentration (electrical percolation threshold). The percolation power law for the conducting regime ($f_{LCMO} > \Phi_c$) returns a critical exponent $t$ value of 2.0±0.2 at room temperature and 2.6±0.2 at 5 K. The authors explain why the difference between the two values of $t$ may result from the contribution of the grain boundaries to the resistivity at low temperatures. On the other hand, the sample porosity of the above medium becomes a significant parameter when the temperature dependence of the thermal conductivity is discussed [15].

One of the related challenges is the optimization of complex multi-phase new materials, beyond just predicting and analysing the existing ones. However, in this paper a simple lattice model serves as a trial approach to gather some information about the percolation behaviour. Interestingly, similar configuration clusters $2 \times 2$ for the square tessellation within real space renormalization group (RSRG) theory were already used to evaluate the effective conductivity of random three-phase composites [16]. However, the utility of the RSRG method depends upon certain underlying assumptions. One of them needs the tessellation to be infinitely large. Then, the position of the percolation threshold on the volume fraction $\phi_H$-axis is independent of the length scale. This is not the approach which we use in this paper.

We restrict ourselves to the non-interacting three phases of a low, moderate and high conductivity, respectively $\sigma_L = 10^{-6}$, $\sigma_M = 1$ and $\sigma_H = 10^6$ in a.u. Using the same basis assumption from Section 2, the straightforward modification of Eq. (2.1) leads to

$$\sigma_x\left(\{n_{m_H, m_L}\}\right) = \frac{k}{l} \sum_{m_H=0}^{k} \sum_{m_L=0}^{k-m_H} \frac{n_{m_H, m_L}}{m_H/\sigma_H + (k - m_H - m_L)/\sigma_M + m_L/\sigma_L} \,. \tag{4.1}$$

Now, further modifications referred to Eqs. (2.3a, b) can be readily derived:

$$P\left(\{n_{m_H, m_L}\}\right) = \frac{l!}{\prod_{m_H=0}^{k} \prod_{m_L=0}^{k-m_H} n_{m_H, m_L}!} \prod_{m_H=0}^{k} \prod_{m_L=0}^{k-m_H} (p_{m_H, m_L})^{n_{m_H, m_L}} \,, \tag{4.2}$$

and

$$p_{m_H, m_L} = \frac{k!}{m_H!(k - m_H - m_L)!m_L!} \phi_H^{m_H} (1 - \phi_H - \phi_L)^{k-m_H-m_L} \phi_L^{m_L} \,. \tag{4.3}$$

Here the notation is an extension of that used in the previous sections. A generalization to the $n$-phase systems is also available.

Now, the main formula (2.4) together with the modified (4.1-3), provides the effective three-phase conductivity $\sigma^*(\phi_H, \phi_L)$ for the two considered extended approaches. The exemplary results are shown for the extended Z4-model in Fig. 4a for $k = 2$ and Fig. 4b for $k = 5$.

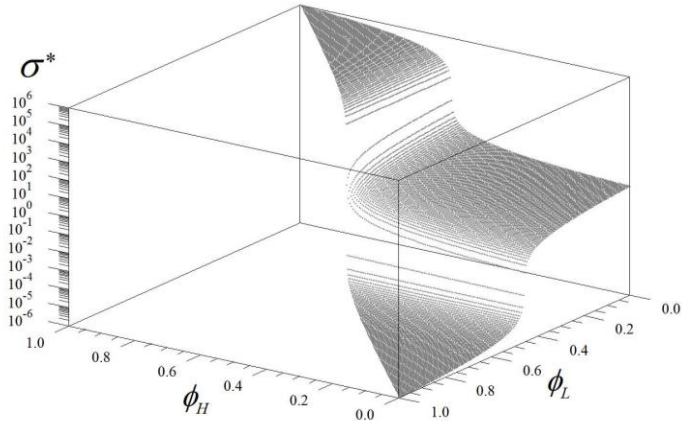

**Fig. 4a.** The effective three-phase conductivity $\sigma^*(\phi_H, \phi_L)$ for the Z4-model and chosen length scale $k = 2$. The conductivity of the $H$, $M$ and $L$-phase components is $10^6$, 1 and $10^{-6}$ in a.u. The characteristic points appear $\sigma^*(0, 1) = \sigma_L$, $\sigma^*(0, 0) = \sigma_M$ and $\sigma^*(1, 0) = \sigma_H$. On the model surface, we observe the percolation behaviour ascribed to each of the above characteristic pairs. Double-threshold routes appear for specified values of the volume fractions $\phi_H$ and $\phi_L$.

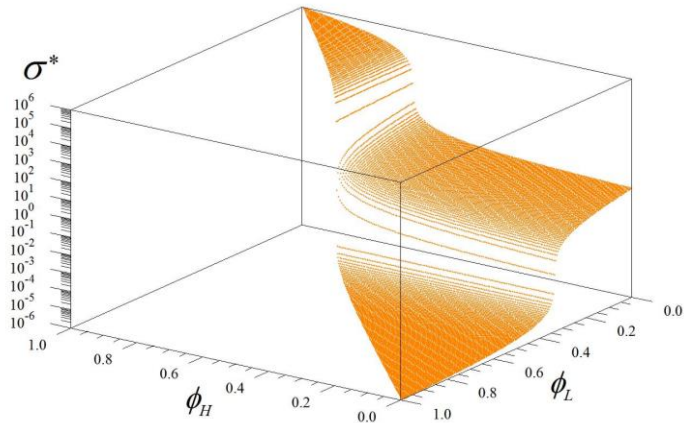

**Fig. 4b.** The same as in Fig. 4a but for the larger length scale $k = 5$.

In turn, Fig. 5a for $k = 2$ and Fig. 5b for $k = 5$ exhibit the corresponding results for the modified Z2-model.

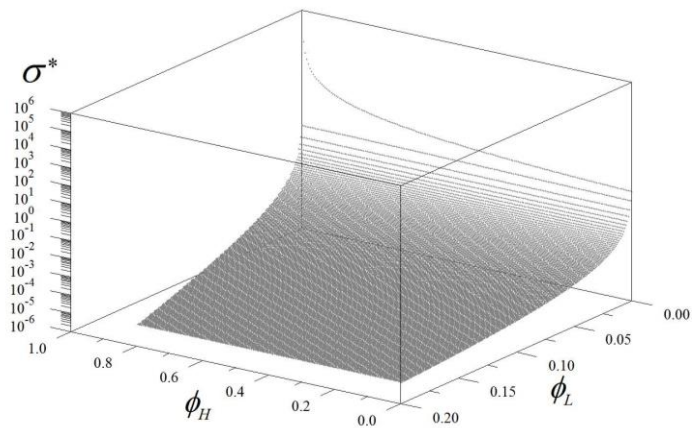

**Fig. 5a.** The same as in Fig. 4a but for the Z2-model. For better visibility, the volume fraction of *L*-phase is cut-off at $\phi_L = 0.2$. Now, the percolation behaviour can be seen in the narrow ranges of high concentrations of appropriate phases, when a phase of higher conductivity becomes the dominant one.

For all the considered models, the characteristic points appear, $\sigma^*(0, 1) = \sigma_L$, $\sigma^*(0, 0) = \sigma_M$ and $\sigma^*(1, 0) = \sigma_H$. One can observe on the depicted effective conductivity surfaces the percolation behaviour ascribed to each of the pairs of points listed above. One can imagine countless double-threshold paths appearing for the specified values of the volume fractions $\phi_H$ and $\phi_L$. However, for the Z2-model the percolation behaviour is clearly seen in the narrow ranges of essentially high volume fractions of the appropriate phases, when a phase of higher conductivity becomes the dominant one. In Figs. 5a, b, the volume fraction of *L*-phase is cut-off at $\phi_L = 0.2$ for better visibility of some details.

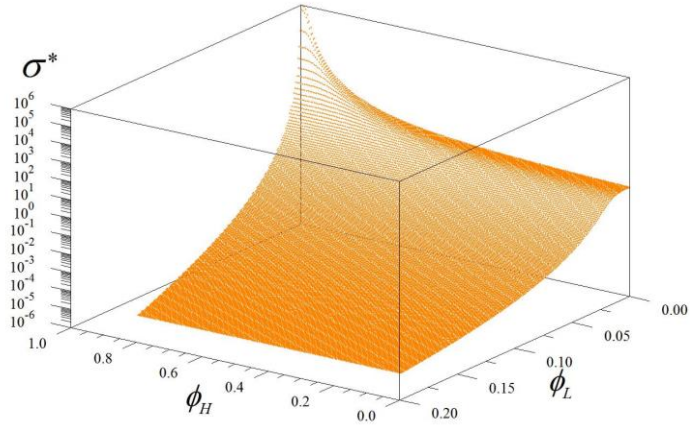

**Fig. 5b.** The same as in Fig. 5a but for the larger length scale $k = 5$.

## 5. Conclusions

In this paper, we investigated the multi-scale percolation behaviour in the high-temperature limit of the specific lattice model that was described in Ref. [8]. We considered basic clusters of $k \times k$ size. At scales available numerically ($k \leq 12$), the multi-scale analysis of the effective conductivity was performed for the EMA-networks with different coordination numbers, i.e. $z = 4$ and $z = 2$. To examine percolation behaviour at any length scale, the simplified variants of models mentioned above were proposed.

We found that the position of the percolation threshold is highly sensitive to the size of a local cluster. However, the slope of the related conductivity curves remains nearly the same. We also observed that the percolation threshold in models with $z = 4$ is a monotonic function of the length scale *k*. On the other hand, in models with $z = 2$ the corresponding function is a non-monotonic one. We argue that the *k*-multiscale displacements of the percolation threshold show a common behaviour, when electrical conductivity is dominated by a few linear conducting paths.

The present approach applies also to multi-phase systems. In the briefly discussed three-phase case, we observed double-threshold paths on the effective conductivity surfaces. The models presented here can be used in further studies, especially at finite temperatures including also particles of various shapes.

**Appendix**

The present extension can also be addressed to a system of interacting particles. However, this point needs a brief clarification. We are concerned about the probably misprinted captions of cases (c) and (d) in Fig. 2 of Ref. [8] related to the probabilities $P(\mathbf{c})$ and $P(\mathbf{d})$ of sets **c** and **d** of possible proper configurations. Each of the configurations of the two sets has appropriate local conductivity along fixed $x$-axis referred to a specified equivalent electric network, $\sigma_x(\mathbf{c})$ for the first set $\mathbf{c} = \left\{ \begin{pmatrix} 1 & 1 \\ 0 & 0 \end{pmatrix}, \begin{pmatrix} 0 & 0 \\ 1 & 1 \end{pmatrix} \right\}$ and $\sigma_x(\mathbf{d})$ for the other one $\mathbf{d} = \left\{ \begin{pmatrix} 1 & 0 \\ 0 & 1 \end{pmatrix}, \begin{pmatrix} 0 & 1 \\ 1 & 0 \end{pmatrix}, \begin{pmatrix} 1 & 0 \\ 1 & 0 \end{pmatrix}, \begin{pmatrix} 0 & 1 \\ 0 & 1 \end{pmatrix} \right\}$ where 1 (0) corresponds to a particle of $H$-phase ($M$-phase). We recommend the following probabilities: $P(\mathbf{c}) = 2\phi^2(1-\phi)^2 e^{\beta(2\mu+\Delta)}/Z$ and $P(\mathbf{d}) = 2\phi^2(1-\phi)^2 e^{2\beta\mu}(1+e^{\beta\Delta})/Z$, where $Z$ is a normalization factor. It should be noticed that in the present paper we consider non-interacting particles only, i.e. we put $\beta = 0$.